\documentclass[12pt,a4paper]{article}
\usepackage[utf8]{inputenc}
\usepackage[T2A,T1]{fontenc}
\usepackage[russian,english]{babel}
\usepackage{booktabs}
\usepackage{graphicx} 
\usepackage{gensymb}
\usepackage{csquotes}
\usepackage[
  style=authoryear,
  sorting=nyt,
  sortcites=true,
  giveninits=true,
  maxbibnames=4,
  minbibnames=3,
  maxcitenames=2,
  mincitenames=1,
  uniquename=false,
  uniquelist=false
]{biblatex}
\usepackage{pdflscape}
\usepackage{indentfirst}

\DeclareBibliographyDriver{article}{%
  \usebibmacro{bibindex}%
  \usebibmacro{begentry}%
  \usebibmacro{author/translator+others}%
  \setunit{\labelnamepunct}\newblock
  \iffieldundef{shortjournal}
  {\printfield{journaltitle}}
  {\printfield{shortjournal}}%
  \setunit*{\space}%
  \iffieldundef{volume}
  {\printfield{number}}
  {\printfield{volume}}%
  \setunit{\addcomma\space}%
  \printfield{pages}%
  \newunit\newblock
  \usebibmacro{addendum+pubstate}%
  \usebibmacro{finentry}%
}

\DeclareFieldFormat[article]{shortjournal}{\mkbibemph{#1}}
\DeclareFieldFormat{pages}{\mkfirstpage{#1}}

\AtEveryBibitem{%
  \clearfield{title}%
  \clearfield{subtitle}%
  \clearfield{shorttitle}%
  \clearfield{month}%
  \clearfield{day}%
  \clearfield{urlyear}%
  \clearfield{urlmonth}%
  \clearfield{urlday}%
  \clearfield{issn}%
  \clearfield{isbn}%
}

\AtBeginBibliography{%
  \setlength{\emergencystretch}{3em}%
  \DeclareNameAlias{default}{family-given}%
  \DeclareNameAlias{sortname}{family-given}%
}

\renewcommand*{\labelnamepunct}{\addspace}
\let\cite\parencite

\DeclareUnicodeCharacter{03B1}{\ensuremath{\alpha}}
\title{\LARGE\bfseries A 2D model of the heliosphere based on SDO coronal hole observations}
\author{%
  \textbf{A.O.~Shiryaev}$^{1,2}$\thanks{Corresponding author: anton.o.shiryaev@gmail.com},
\textbf{Yu.S.~Shugay}$^{2}$,
\textbf{K.B.~Kaportseva}$^{2}$.}
\date{\small\itshape
  $^{1}$Faculty of Space Research, M.V.~Lomonosov Moscow State University,\\
  Leninskie Gory~1, Building~52, Moscow, 119234 Russia\\[0.5em]
  $^{2}$Space Physics Laboratory, Skobeltsyn Institute of Nuclear Physics,\\
M.V.~Lomonosov Moscow State University, Leninskie Gory, GSP-1, Moscow, 119991 Russia}

\begin{document}

\maketitle

\begin{abstract}
  We present SWX-2D, a two-dimensional extension of an empirical coronal hole-based solar wind forecast.
  It prescribes radially outward solar wind velocities at Carrington longitudes using speeds inferred from Earth-facing SDO observations and ballistically propagates solar wind parcels through the ecliptic plane to 1~AU.
  The model was evaluated over the interval from January~1,~2018, to January~1,~2019, against hourly in situ ACE and STEREO\mbox{-}A measurements.
  With the empirical slow wind correction, SWX-2D and the operational one-dimensional SWX forecast agreed with \(r=0.93\) and an RMSE of \(30~\mathrm{km\,s^{-1}}\), confirming that the forecast mechanism was largely preserved.
  In situ validation of the raw output yielded \(r=0.55\) at ACE and \(r=0.40\) at STEREO\mbox{-}A, as compared to \(r=0.44\) and \(r=0.63\), respectively, for 27.27-day recurrent forecasts evaluated on the same timestamps.
  ICME exclusion did not materially change the aggregate correlations.
  The results support extending the forecast away from the Earth--Sun line, while identifying limitations from aged Earth-view observations and simplified propagation.
\end{abstract}


\section{Introduction}
\subsection{Motivation}
Quasistationary solar wind (SW) streams flowing from the Sun's corona consist of a fast and a slow component, which are characterized by different densities, temperatures, ion compositions and solar sources \cite{Feldman1976, Feldman2005}.
The fast component is formed by high-speed streams (HSSs) associated with coronal holes (CHs) \cite{Krieger1973}, while the slow component is associated with several coronal source regions, including the streamer belt and boundaries of open magnetic flux \cite{Feldman2005, Crooker2012}.
HSSs interact with slow solar wind that precedes them to form stream interaction regions (SIRs), also called corotating interaction regions (CIRs) when they recur and are associated with a persistent coronal source \cite{Richardson2018}.
High-speed streams and CIRs can by themselves cause recurring geomagnetic disturbances \cite{Tsurutani2006}.
They also contribute to the quasistationary background through which interplanetary coronal mass ejections (ICMEs) propagate \cite{Gopalswamy2009, Heinemann2019}.
Interactions between ICMEs and HSSs can significantly increase ICME geoeffectiveness \cite{Heinemann2024}.
These effects motivate forecasting the ambient solar wind both as a driver of recurrent geomagnetic disturbances and as the background through which ICMEs propagate.
Extending such forecasts beyond the Earth--Sun line would also enable their evaluation at other heliolongitudes and provide a spatial description of the heliospheric background.
Furthermore, empirical CME propagation models combining the drag-based model (DBM) with cone geometry allow propagation directions away from the Earth--Sun line \cite{Zic2015, Dumbovic2021a} and could therefore benefit from ambient solar wind speed estimates along those paths.

\subsection{Modeling}
The Alfv{\'e}n surface is commonly used to mark the transition between the solar corona and the heliosphere \cite{DeForest2014}.
It is defined dynamically as the surface beyond which the solar wind becomes super-Alfv{\'e}nic and information can no longer propagate upstream into the corona.
A common approach to heliospheric solar wind modeling is to estimate the solar wind speed at an inner boundary that roughly corresponds to this surface and then propagate this boundary condition outward through the heliosphere. In physics-based models, this boundary condition is usually derived from coronal magnetic field models.
For example, the Wang-Sheeley-Arge (WSA) model estimates the solar wind speed from the geometry of the coronal magnetic field, while the ENLIL magnetohydrodynamic (MHD) model propagates the resulting solar wind through the inner heliosphere \cite{SheeleyJr.2017, Odstrcil2003}.
WSA-ENLIL and other magnetohydrodynamic models \cite{Arutyunyan2025} provide a physically detailed description of the heliospheric plasma, but they require substantially more input information and computational resources than empirical or kinematic approaches.

Reduced heliospheric propagation models occupy an intermediate position between empirical point forecasts and full MHD simulations. Ballistic models propagate solar wind parcels radially with prescribed speeds, while HUX-type models \cite{Riley2011, Barnard2022} approximate the longitudinal evolution of solar wind speed using simplified hydrodynamic assumptions.
Such models are computationally efficient and useful for reconstructing or forecasting large-scale solar wind speed structure, although they do not explicitly model density, magnetic field evolution, or the full physics of stream interaction regions.

Coronal holes (CH) are low-density coronal regions generally associated with predominantly open magnetic flux \cite{Zhang2003}.
CH are often associated with high-speed streams \cite{Krieger1973, Rotter2012}.
They appear dark in extreme ultra-violet images and are commonly detected in images from the AIA instrument aboard the SDO spacecraft \cite{Lemen2012, Reiss2024}.
Coronal hole area on the visible solar disk correlates with observed solar wind speed \cite{Rotter2012, Shugai2021}, which makes it possible to use CH observations for solar wind speed forecasting.

Coronal hole area and position can be used to predict solar wind speed \cite{Vrsnak2007}.
The Empirical Solar Wind Forecast (ESWF) uses coronal hole area to predict SW speed at Earth, adjusting the parameters of the underlying CH area-SW speed relation based on the three previous rotations \cite{Rotter2015}, and yields a root mean square error (RMSE) comparable to WSA \cite{Reiss2016}.

A similar empirical forecasting framework described by \textcite{Shugay2011} is the foundation of the operational SWX solar wind forecast provided by the Space Weather eXchange of SINP MSU (\url{https://swx.sinp.msu.ru/models/solar_wind.php}).
In this model, Earth-facing CH area is converted to solar wind speed.
Predicted arrival times near ACE/L1 are obtained by adding the ballistic travel time over a fixed distance of approximately 1 AU to the SDO observation time.
Overtaking is treated kinematically: when a later, faster parcel is predicted to arrive before earlier, slower parcels, it supersedes their forecast values.
The resulting arrival-time series is then interpolated.
This prescription permits stream overtaking without resolving the associated compression and rarefaction regions.
The operational implementation of this model, referred to as simply SWX hereafter, and its validation over 2010--2019 are described in \textcite{Shugai2021}.

Existing CH area forecasting models primarily produce a one-dimensional time series at or near Earth.
In this work, SWX is extended from a one-point forecast at L1 to SWX-2D, which assigns radially outward boundary velocities, with magnitudes derived from CH observations, to Carrington longitudes and ballistically propagates solar wind parcels through the ecliptic plane to 1~AU, allowing the same reconstructed solar wind structure to be sampled at multiple radii and heliolongitudes.
Its kinematic treatment captures stream overtaking while omitting the plasma and magnetic pressure dynamics represented in MHD models.

\section{Methods}
\subsection{Using SDO images to obtain solar wind speed at the inner boundary}
The starting point of the modeling of quasistationary solar wind in the presented model is the determination of coronal hole boundaries in SDO images, primarily from the 193~\AA{} channel, which contains Fe~XII and Fe~XXIV emission lines associated with the solar corona \cite{Lemen2012}.
Coronal hole emission being lower owing to reduced plasma density and, to a lesser extent, lower temperature, CHs appear dark in images from this channel \cite{Hahn2011}.

To match the total SW travel distance to that of the SWX model the inner boundary of the model, at which the propagation starts, is formally set at 1 AU from the SDO spacecraft, the data from which is used to calculate the velocities.
While SWX only prescribed SW speed at 1 AU and SWX-2D extends that to a 2D grid, we obviously do not consider SWX-2D a coronal model; the region within 20 solar radii, corresponding roughly to the Alfv{\'e}n boundary, is therefore occluded on visualizations.

\begin{figure}
  \centering
  \includegraphics[width=0.6\textwidth]{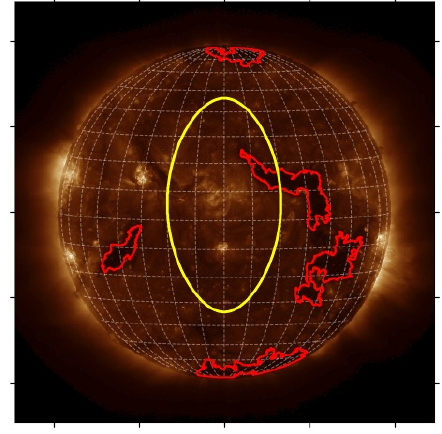}
  \caption{An example of the operation of the coronal hole detection algorithm on an SDO/AIA 193~\AA{} image. Areas classified as coronal holes by the threshold algorithm are shown in red, and the oval bounding the area used to calculate solar wind speed is shown in yellow.}
  \label{fig:cd_detection}
\end{figure}

A simple threshold algorithm is used to determine coronal hole boundaries.
Parts of the image that are darker than the mean intensity of the solar disk scaled by an empirically adapted coefficient (0.25 for images before Jan~1,~2019 and 0.3 after) are classified as coronal holes.
Only coronal hole area that falls inside an oval spanning latitudes of $\pm40\degree$ and longitudes of $\pm20\degree$ is considered relevant for Earth-facing high-speed stream forecasting; the ratio of cumulative CH area inside that oval to the area of the entire visible solar disk is used to determine solar wind speed at the inner boundary via the following empirical relationship:
\begin{equation}
\label{eq:closure}
  v(S) = v_{min} + A \ S^\alpha,
\end{equation}
where $S$ is the CH area ratio expressed as a percentage, $v_{min}=300~\mathrm{km\,s^{-1}}$ is a constant speed representing slow solar wind.
The operational SWX coefficients $A=210$ and $\alpha=0.4$, obtained by least-squares minimization before the modeled interval, remained fixed and were used operationally through December~31,~2018. Beginning January~1,~2019, the operational model used the subsequently recalibrated coefficients $A=180$ and $\alpha=0.6$ \cite{Shugai2021}.
An example of the threshold algorithm and the oval containing the coronal hole area used for the Earth-directed forecast is shown in Figure~\ref{fig:cd_detection}.
Each solar wind `packet' is assumed to propagate with a constant velocity: its speed remains constant and its direction remains radially outward along its propagation ray.
When a faster solar wind packet overtakes a slower one, meaning both an earlier slower packet and a later faster packet exist in some modeling cell at the same time, the faster packet replaces the slower one.
The coefficients of the empirical SW model were optimized while accounting for this overtaking mechanism by comparing fully propagated and interpolated SW to ACE data \cite{McComas2022}.

\subsection{Two-dimensional modeling}
\label{sec:2d}
SWX is one-dimensional and describes SW propagation only along the Earth--Sun line \cite{Shugai2021}.
SWX-2D extends this approach to model a slice of the ecliptic plane heliosphere extending to 1~AU.
To do so, we assume that the coronal holes observed when SDO was facing a certain Carrington longitude, and the SW speed calculated from their area, remain essentially unchanged until that longitude is observed again one Carrington rotation later.
This assumption is not unique to SWX-2D; both recurrence-based approaches \cite{Owens2013} and models based on synoptic maps of the Sun, such as WSA-ENLIL \cite{Wang1990, Arge2000, SheeleyJr.2017}, are based on data from vantage points along the Earth--Sun line only.
Furthermore, coronal holes associated with geoeffective HSS often exist for more than one Carrington rotation (CR) \cite{Heinemann2020}, which is the principle behind recurrent forecasting.

Modeling is performed on a discrete polar coordinate grid.
We calculate the SW speed associated with Carrington longitude from the CH area observed when that longitude was facing the Earth.
That longitude continuously emits solar wind with that speed until it is observed again 1~CR later.
Emitted solar wind is ballistically propagated from the inner boundary out to 1~AU.
Faster solar wind streams can overtake slower ones emitted earlier; when a solar wind packet is written to a cell that already contains an earlier packet it overwrites it after all propagation has been completed; in other words we assume that the solar wind speed in each modeling cell at each point in time is the fastest solar wind speed that reached that coordinate at that time.
The angular resolution of SWX-2D corresponds to two hours of the Sun's rotation, or $\approx1.1\degree$; this resolution is comparable to the angular resolution of other models (HUXt: azimuthal $2.8\degree$, radial $1.5 R_{solar}$ \cite{James2023}; WSA-ENLIL: azimuthal $4\degree$ \cite{Dewey2016}).

As in SWX, hourly SDO images are the primary input.
The ACE and STEREO\mbox{-}A time series are also analyzed at hourly resolution.

The nominal radial resolution is $1.5 R_{solar}$, the same as that used by HUXt \cite{James2023} and WSA-ENLIL in the inner heliosphere.
The radial grid extends from $0$ to $218 R_{solar}$, slightly beyond 1~AU, to include the ACE position throughout the validation interval.
All radial intervals have a width of $1.5 R_{solar}$ except for a single shorter innermost interval from $0$ to $0.5 R_{solar}$.
To avoid fast solar wind leap-frogging cells, the temporal step and radial spacing need to satisfy $v_{max} \cdot \Delta t \le \Delta r$ for every radial interval.
With an hourly step, the corresponding speed limit is approximately $290~\mathrm{km\,s^{-1}}$ for the nominal intervals and $97~\mathrm{km\,s^{-1}}$ for the innermost interval; thus, practically any modeled solar wind can skip cells when hourly temporal resolution is used.
A stream traveling at $700~\mathrm{km\,s^{-1}}$ advects by $\approx3.6 R_{solar}$ in an hour; accommodating this displacement would require an unreasonably coarse radial resolution.

There are three core problems that arise from advecting parcels over a discrete grid: the occurrence of empty cells in rarefaction regions, the need to account for stream overtaking in compression regions, and physically unjustified interleaving of slow and fast streams due to cell skipping.
When a slow solar wind stream occurs after a faster one, a rarefaction arises.
Since SWX-2D does not account for any azimuthal stream interaction or expansion, and no solar wind packet launched from the Sun reaches these cells, they remain empty.
In the opposite situation, a faster solar wind stream follows and overtakes a slower one, producing a compression region.
Since the ballistic propagation model is insufficient to account for interplanetary shocks, we account for such overtaking by taking the fastest solar wind stream that has reached a particular simulation cell at a particular moment, to preserve the continuity of the faster streams that are more relevant for applied space weather tasks; in other words, when overtaking/collision occurs, the parcels of the faster stream merely `overwrite' the slower ones.
We also investigated averaging the speeds of parcels that occupied a particular cell at the same moment, but that approach led to a much worse representation of the CIR interface, smearing the sharp edge.
Interleaving occurs when a fast solar wind catches up with and overtakes a slower one, but due to the higher speed of its constituent parcels they skip some cells that were not skipped by parcels of the slower stream being overtaken.
This effect appears as a sequence of alternating low and high speeds in a time series plot at a particular point on the simulation grid.

\begin{figure}
  \centering
  \includegraphics[width=\textwidth]{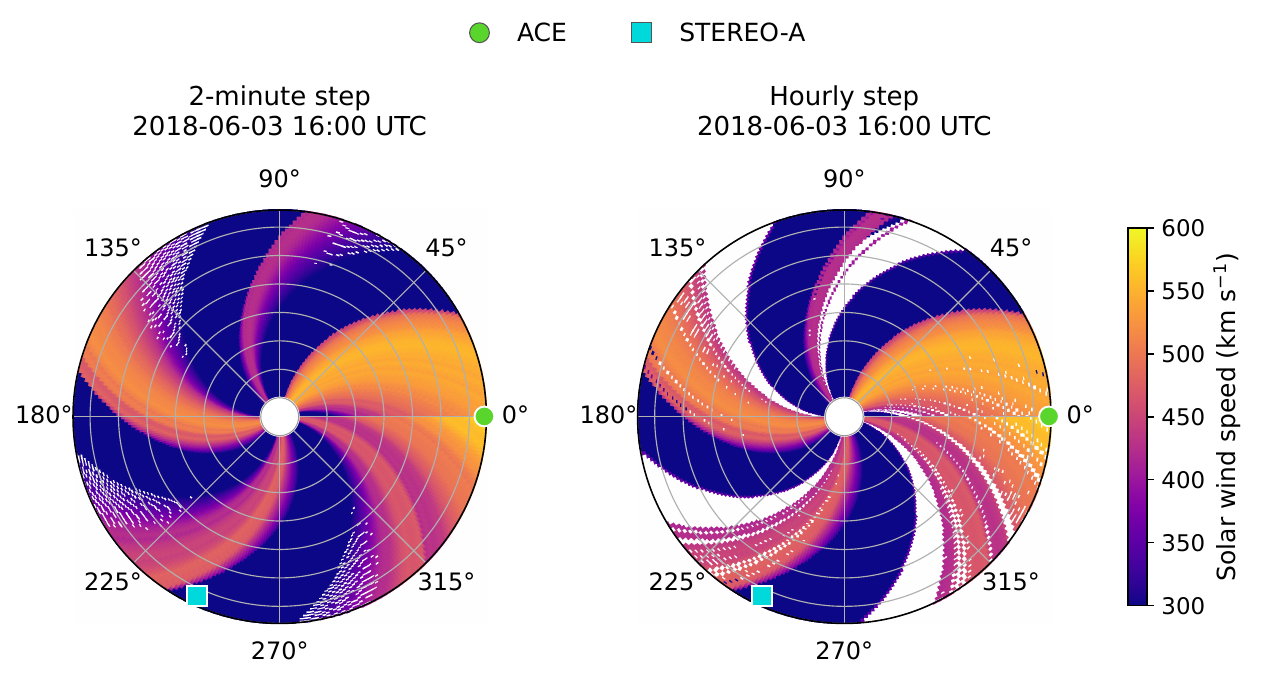}
  \caption{Comparison of the SWX-2D `fill rate' with hourly temporal resolution on the right and 2-minute resolution on the left.}
  \label{fig:interpolation}
\end{figure}

The empty cell problem can be overcome via `r-shell' interpolation, e.g., interpolation of SW speed between cells with different azimuthal coordinates but matching radial and temporal coordinates.
But such a solution will smear interleaved slow and fast streams into a single stream, and the speed of the resulting stream will inch closer to the speed of the faster solar wind stream as resolution increases.
Interleaving could be solved by depositing a parcel's SW speed value into each cell between $r(t_i)$ and $r(t_{i+1})$ as it travels between them in a single simulation step.
A simpler solution to both problems is increasing the temporal resolution independently of the azimuthal resolution by interpolating the emitted solar wind speed at the SWX-2D inner boundary.
With an internal temporal resolution of $2$~minutes, a parcel traveling at $700~\mathrm{km\,s^{-1}}$ advects by approximately $0.12 R_{solar}$, less than even the minimum radial spacing of $0.5 R_{solar}$.
This prevents radial cell skipping and the associated interleaving at the modeled speeds.
Increasing the internal temporal resolution also means that the radial gaps between discrete parcels become smaller than the grid spacing, so the fixed grid is much less likely to contain cells with no assigned solar wind speed (Figure~\ref{fig:interpolation}).

In SWX, interpolation was performed on the time series at ACE, whereas in SWX-2D it is applied to data at the longitudinal boundary condition before propagation.
This change is not neutral, since the ballistic travel time depends on solar wind speed, $\tau(v)=L/v$.
During a transition to higher speed, later and faster parcels have shorter travel times and therefore arrive closer together, steepening the leading edge of the high-speed stream.
During the subsequent transition to lower speed, later and slower parcels lag progressively further behind, stretching the trailing edge.

\subsection{Validation}
\label{sec:validation_setup}
Validation is performed against measurements from two spacecraft: STEREO\mbox{-}A and ACE.
Validation against ACE data allows us to ensure continuity of SWX-2D with its one-dimensional predecessor, SWX.

The period used for validation in this paper was selected from the late declining phase of Solar Cycle~24.
Since it is characterized by low-latitude coronal hole activity \cite{Luhmann2009} and reduced active region activity and coronal mass ejection (CME) occurrence, it is well suited to testing SWX-2D, which relies exclusively on CH area data and does not currently account for ICMEs.
It is important to note that STEREO\mbox{-}A moved from approximately $120\degree$ to $100\degree$ west of Earth during 2018.
The Earth-view observations used to construct the boundary condition were therefore approximately 18--20~days old when radial propagation began and, after a further 2.8--5.6~days of solar wind propagation from the inner boundary, approximately 21--26~days old when the solar wind reached STEREO\mbox{-}A.
This observation age may reduce forecast accuracy.

SWX-2D does not account for ICMEs and their interaction with ambient solar wind, nor does it account for variability of the slow solar wind.
We analyzed model metrics for two data sets: all data and data with ICME intervals excluded.
ICME exclusion for the ACE data was performed using data from the Richardson \& Cane catalog \cite{Richardson1995}, the SRI RAS Large-Scale Solar Wind Event catalog \cite{Yermolaev2009} and events from the CCMC CME Scoreboard \cite{Riley2018}, as presented by the SINP MSU unified ICME list \cite{Shiryaev2024}.
A total of 39 ICMEs were excluded, 29 of which were only present in the SRI RAS catalog.
All 6 ICME intervals observed at STEREO\mbox{-}A and listed in the ICMECat catalog \cite{Mostl2017,Mostl2020} were excluded from data for that satellite.
To minimize contamination by ICME-associated disturbances, including preceding shocks and sheaths and the disturbed solar wind following each event, the catalogued ICME intervals were conservatively extended by 12 hours at both ends.
This approach accounts for uncertainties in the catalogued event boundaries at the cost of potentially excluding some undisturbed solar wind.

Validation is performed on observation and hindcast data with hourly resolution.

\section{Results}
The methods described in this study were used to model the heliosphere to 1~AU over the period from January~1,~2018, to January~1,~2019.

To balance computational cost against grid cell coverage, an internal temporal resolution of 2~minutes has been selected.
The 2-minute resolution is not interpreted as the physical or forecast accuracy of SWX-2D at that timescale.
It is used solely as an internal numerical parameter to provide more uniform coverage of the computational grid during solar wind propagation, as detailed in Section~\ref{sec:2d}.
The output time series and the subsequent validation are therefore considered at hourly resolution, consistent with the resolution of the input data.

\begin{figure}
  \centering
  \includegraphics[width=\textwidth]{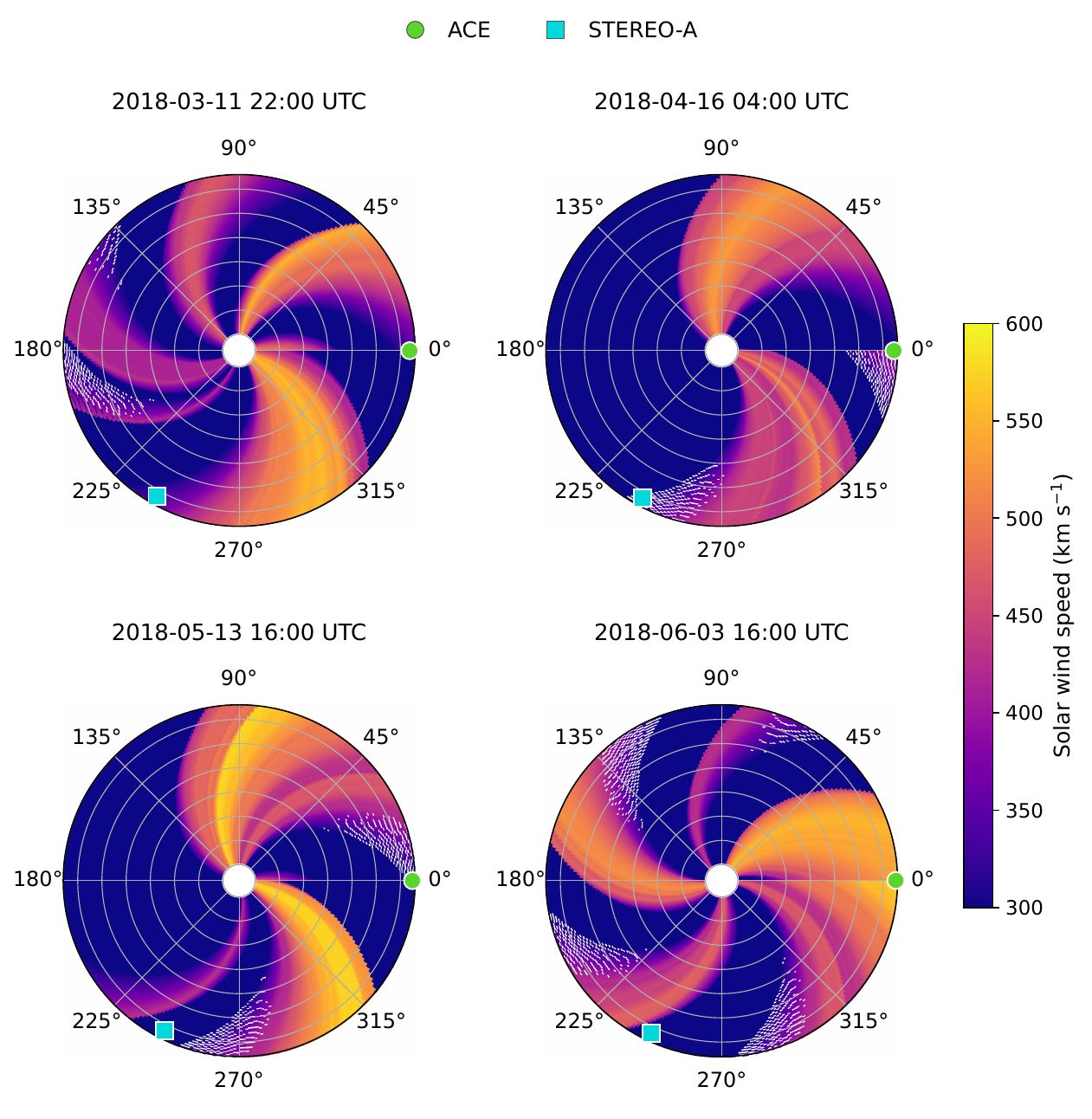}
  \caption{Heliosphere modeling results at several points in~2018.}
  \label{fig:heliosphere}
\end{figure}

Several example images of the simulated heliosphere are shown in Figure~\ref{fig:heliosphere}.

\begin{figure}
  \centering
  \includegraphics[width=\textwidth]{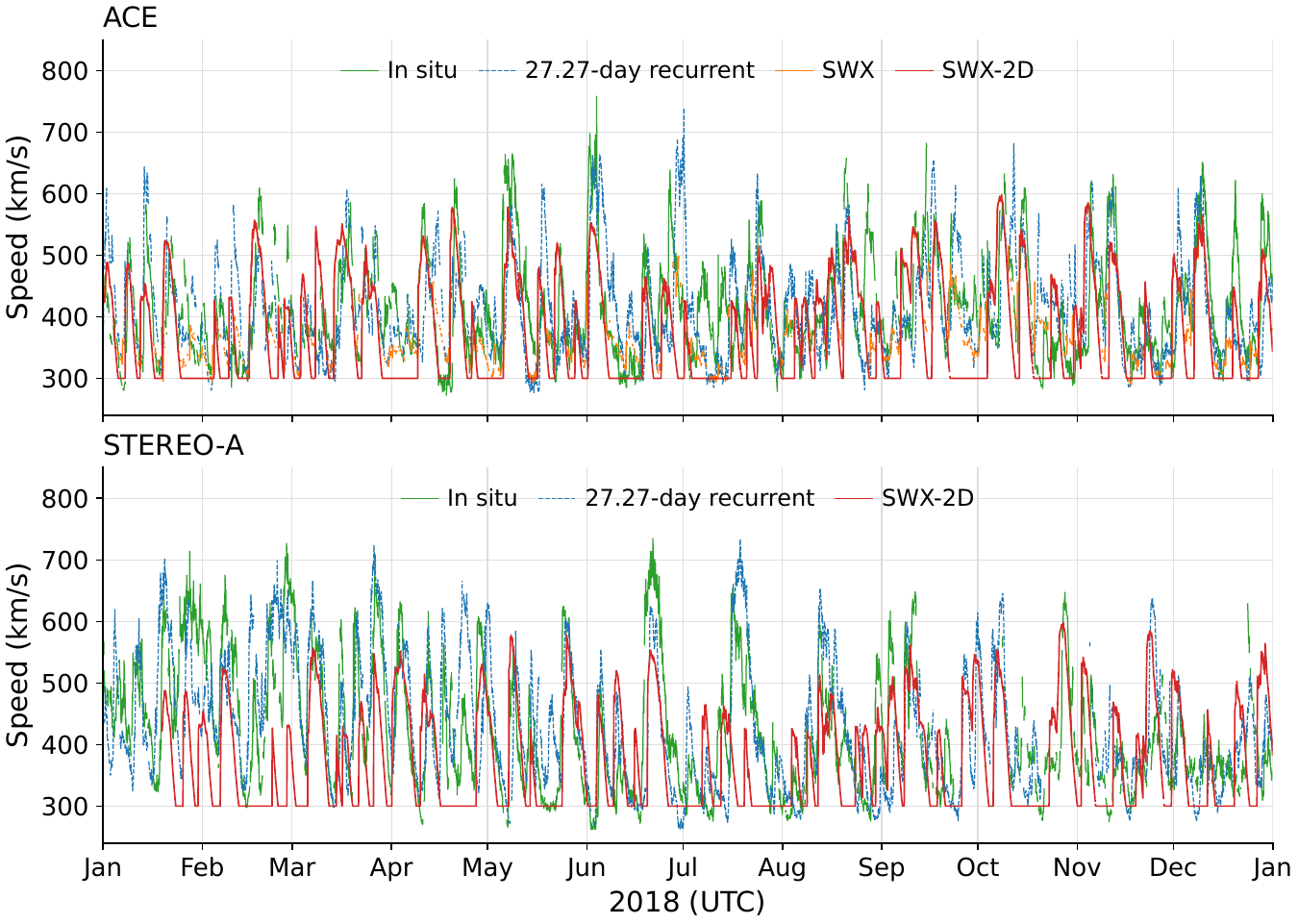}
  \caption{Hourly solar wind speed profiles for~2018 at ACE (upper panel) and STEREO\mbox{-}A (lower panel).
    The SWX-2D prediction with no slow SW correction applied is shown in red, in situ measurements in green, and the 27.27-day recurrent forecast in dashed blue; the one-dimensional SWX forecast, shown in orange, is included only in the upper panel.
  Gaps denote intervals for which the corresponding plotted series is unavailable.}
  \label{fig:sw_time_series}
\end{figure}

The hourly solar wind speed time series used for the comparisons below are shown in Figure~\ref{fig:sw_time_series}.
The upper panel compares the raw SWX-2D prediction at ACE with native ACE measurements, the one-dimensional SWX forecast, and the recurrent forecast.
The lower panel shows the corresponding raw SWX-2D, in situ, and recurrent time series at STEREO\mbox{-}A.
The SWX-2D series at STEREO\mbox{-}A begins on January~19,~2018, when model output first becomes available at the spacecraft position.

\clearpage
\subsection{Comparison of SWX-2D with SWX}
\label{sec:validation_earth}
The development of SWX-2D required modifications to the interpolation architecture and model resolution described in Section~\ref{sec:2d}.
To verify the validity of the new interpolation architecture, we compared SWX-2D with SWX.
For this comparison, we applied the empirical slow SW model \cite{Shugai2021} to the SWX-2D time series at ACE:
\begin{equation}
\label{eq:slow}
  v_{\mathrm{slow}}(t)
  =
  \frac{1}{2}
  \left[
    v_{\mathrm{ACE}}(t-672\,\mathrm{h})
    +
    \min_{\tau \in [t-336\,\mathrm{h},\,t-1\,\mathrm{h}]}
    v_{\mathrm{ACE}}(\tau)
  \right].
\end{equation}
Here, $v_{\mathrm{ACE}}$ is the ACE solar wind speed series resampled to hourly means, and the minimum is taken over its hourly samples within the stated interval.
Missing values in this input series are filled by time interpolation, followed by forward and backward filling of any remaining gaps.
In the SWX-2D time series at ACE, like in the 1D SWX forecast, the empirical slow SW model is applied post hoc by replacing the constant $300~\mathrm{km\,s^{-1}}$ values produced when no CH is observed with $v_{\mathrm{slow}}(t)$.

During the interval from January~1,~2018, to January~1,~2019, SWX-2D with the empirical slow SW model and SWX had $N=4279$ paired hourly values.
Their Pearson correlation coefficient was $r=0.93$, with $\mathrm{RMSE}=30~\mathrm{km\,s^{-1}}$ relative to SWX.

The resulting metrics indicate close consistency between SWX-2D with the empirical slow SW model and SWX after the transition to the new interpolation and propagation scheme.

Because SWX-2D and SWX share both their empirical input relationship and the empirical slow SW model, this comparison is a consistency test rather than an independent validation of forecast skill.
Independent validation against in situ measurements is considered separately.

\subsection{Comparison of SWX-2D with in situ measurements}
\label{sec:2d_in_situ}

SWX-2D was validated against hourly solar wind speed measurements from ACE and STEREO\mbox{-}A.
The comparison used SWX-2D output without the empirical slow SW model, because the empirical slow SW model is validated only at ACE and is not applicable to the STEREO\mbox{-}A forecast.
A 27.27-day recurrent forecast, constructed from measurements at the corresponding spacecraft and with gaps of up to six hours interpolated, was used as a reference baseline.
\textcite{Owens2013} found that 27-day persistence was competitive with physics-based forecasts, including WSA-ENLIL, during solar minimum.
\textcite{Reiss2016} found that ESWF and WSA were broadly comparable to 27-day persistence over 2011--2014.

\begin{figure}[p]
  \centering
  \includegraphics[width=0.95\textwidth]{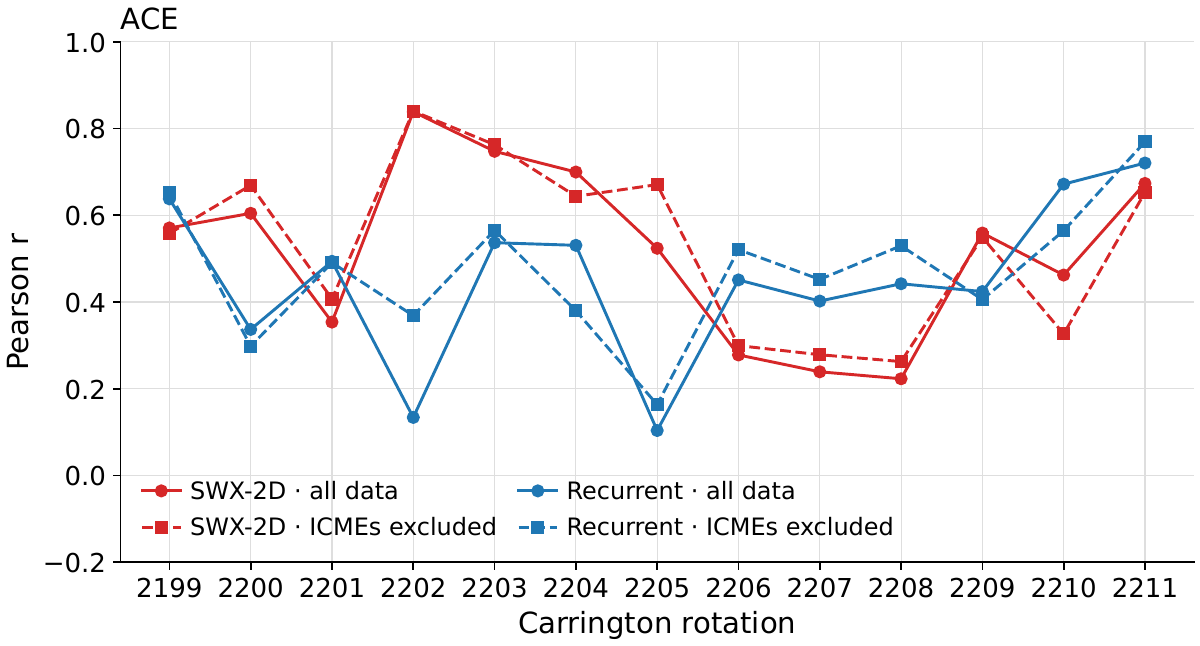}
  \includegraphics[width=0.95\textwidth]{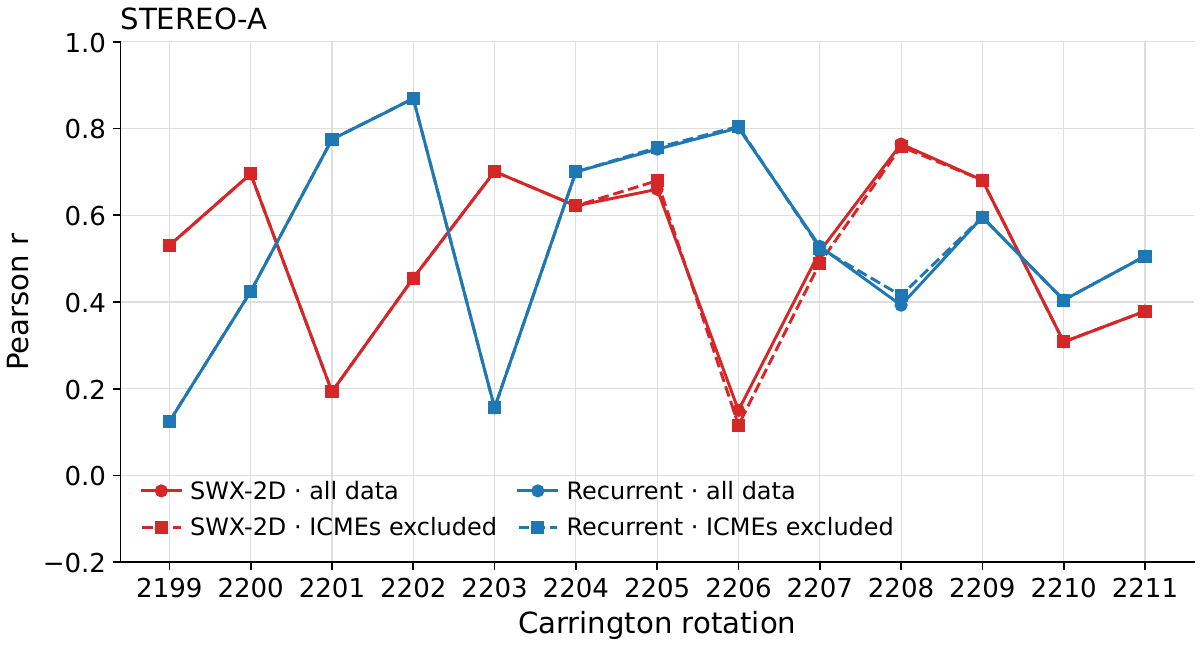}
  \caption{Pearson correlation coefficients, calculated separately for each Carrington rotation for SWX-2D and the recurrent forecast, at ACE (upper figure) and STEREO\mbox{-}A (lower figure).
  Solid lines with circles show the results for all available solar wind data, while dashed lines with squares show the results after excluding ICME intervals padded by 12 hours at both ends.}
  \label{fig:2d_in_situ_cr}
\end{figure}

The metrics were calculated both for all available observations and after excluding the ICME intervals defined in Section~\ref{sec:validation_setup}.
Missing in situ measurements were not interpolated.
For each spacecraft, both forecasts were evaluated on the common timestamps for which the observation, SWX-2D forecast, and recurrent forecast were all available.
The full-period common sample sizes, correlations, and RMSE values for both forecasts at each spacecraft are summarized in Table~\ref{tab:2d_in_situ}.
The correlations calculated separately for each Carrington rotation are shown in Figure~\ref{fig:2d_in_situ_cr}.
CR~2212 was excluded from the per-rotation analysis because only 42\% of that rotation falls within the validation interval; its samples remain included in all full-period metrics.

\begin{table}[h]
  \centering
  \caption{Comparison of SWX-2D and the recurrent forecast to in situ solar wind speed measurements from January~1,~2018, to January~1,~2019.
    Metrics are reported for all available data without excluding ICME intervals.
    ``SWX-2D'' denotes the forecast without the empirical slow SW model, while ``Recurrent'' denotes the 27.27-day recurrent forecast.
    $N$ is the common number of hourly samples used for both forecasts at each spacecraft, and $r$ is the Pearson correlation coefficient between each forecast and the corresponding observations.
  RMSE is the root mean square error and is expressed in $\mathrm{km\,s^{-1}}$.}
  \label{tab:2d_in_situ}
  \begin{tabular}{llrrr}
    \hline
    Spacecraft & Forecast & $N$ & $r$ & RMSE \\
    \hline
    ACE      & SWX-2D             & 6155 & 0.55 &  89 \\
    ACE      & Recurrent          & 6155 & 0.44 &  88 \\
    \hline
    STEREO\mbox{-}A & SWX-2D             & 5628 & 0.40 & 113 \\
    STEREO\mbox{-}A & Recurrent          & 5628 & 0.63 &  89 \\
    \hline
  \end{tabular}
\end{table}

As shown in Table~\ref{tab:2d_in_situ}, on the common all-data support at ACE, SWX-2D produced a correlation coefficient of $r=0.55$, compared with $r=0.44$ for the recurrent forecast.
After exclusion of the 39 ICMEs present in the unified catalog~(\ref{sec:validation_setup}), the common support contained $N=4907$ hourly samples; the correlations were $r=0.55$ for SWX-2D and $r=0.46$ for the recurrent forecast, with RMSE values of $91$ and $85~\mathrm{km\,s^{-1}}$, respectively.
SWX-2D therefore had the higher correlation at ACE, while its RMSE was larger than that of the recurrent forecast both for all data and after ICME exclusion.

To provide a simple recurrence-based reference for the similarity of the solar wind observed at ACE and STEREO\mbox{-}A, the ACE time series was shifted by a fixed delta of 19~days, chosen as the rounded midpoint of the independently estimated 18--20-day recurrence delay and not optimized against the measured solar wind speeds.
Thus, at each STEREO\mbox{-}A timestamp $t$, the ACE measurement at $t-19~\mathrm{days}$ was used.
On the $N=5439$ hourly timestamps for which the shifted ACE measurements, STEREO\mbox{-}A measurements, and two-dimensional STEREO\mbox{-}A forecast were all available, the ACE\mbox{--}STEREO\mbox{-}A correlation was only $r=0.28$, whereas the model\mbox{--}STEREO\mbox{-}A correlation on the same support was $r=0.39$, and was also $r=0.39$ after exclusion of 6 ICMEs from ICMECat.
The SWX-2D correlation varied substantially among Carrington rotations: it exceeded $0.5$ in eight of the thirteen CR intervals retained for per-rotation analysis and reached $r=0.76$ in CR~2208 and $r=0.66$ in CR~2205.

On the common STEREO\mbox{-}A support after ICME exclusion ($N=5513$), the correlations were $r=0.39$ for SWX-2D and $r=0.63$ for the recurrent forecast, with RMSE values of $114$ and $89~\mathrm{km\,s^{-1}}$, respectively.
The recurrent forecast therefore showed substantially stronger agreement with the STEREO\mbox{-}A observations both before and after ICME exclusion, as well as lower RMSE.

Excluding ICMEs produced no material improvement in SWX-2D at either spacecraft.
At ACE, this exclusion removed approximately 20.3\% of the common observations, with little change in correlation.
At STEREO\mbox{-}A, only approximately 2.0\% of the common observations fell within the padded ICME exclusion intervals, and all metrics remained effectively unchanged.
At the level of individual Carrington rotations, however, long cataloged ICME occupancy sometimes coincided with low correlation even after the ICME intervals were excluded.
At ACE, CRs~2206--2208 retained correlations of $r=0.30$, $0.28$, and $0.26$, respectively; at STEREO\mbox{-}A, CR~2206 had the largest cataloged ICME duration and a post-exclusion correlation of only $r=0.12$.
Higher cataloged ICME occupancy tended to coincide with lower post-exclusion correlation at ACE, whereas no clear monotonic relationship was evident at STEREO\mbox{-}A; these descriptive associations do not establish causality.

\section{Discussion}
\label{sec:discussion}
Both the CH\mbox{-}SW closure (equation~\ref{eq:closure}) and the empirical slow SW model were originally calibrated using the SWX interpolation and propagation scheme and ACE data.
A natural next step is to recalibrate the CH-SW closure using observations from additional spacecraft, while reassessing the ACE-specific slow SW model separately.
Because STEREO\mbox{-}A made its Earth flyby in August~2023, a hindcast targeting late 2023 or 2024 would provide a test under substantially more favorable longitudinal geometry.
The Parker Solar Probe and Solar Orbiter could also be used during parts of their trajectories that are relatively close to the solar ecliptic plane.

The weak fixed-delay ACE--STEREO\mbox{-}A correlation reported in Section~\ref{sec:2d_in_situ} is consistent with substantial evolution of solar wind structures between their encounters with the two spacecraft.
That SWX-2D nevertheless achieved nontrivial correlations during several Carrington rotations suggests that the coronal hole-derived SW speed remains somewhat predictive away from the Earth--Sun line.
The remaining disagreement may also be affected by the heliographic latitude separation of ACE and STEREO\mbox{-}A.
Over the 2018 validation interval, their median absolute heliographic latitude separation was \(8.6\degree\), compared to \(0.25\degree\) during June--November~2023 and \(0.18\degree\) at the crossing of the Earth--Sun line on August~12,~2023.
Previous work \cite{Shugai2009} showed that shifting the coronal hole sampling region by \(10\degree\) in heliographic latitude could substantially change the correlation between coronal hole area and solar wind speed during the 2007 solar minimum.
This indicates that spacecraft at different heliographic latitudes may sample different parts of a high-speed stream, particularly when the stream originates from an equatorward extension of a polar coronal hole.
Comparison during the 2023 STEREO\mbox{-}A Earth flyby could test the effect of reduced spacecraft separation, although it would not isolate the contribution of any single coordinate and would preserve the small radial separation between ACE and STEREO\mbox{-}A.

The lower full-year correlation at STEREO\mbox{-}A is consistent with the age of the Earth-view observations used to construct its boundary condition, quantified in Section~\ref{sec:validation_setup}, which remains a principal limitation of the approach.
This limitation could also be mitigated by the use of synthetic synoptic maps that use neural networks to predict the evolution of the photospheric magnetic field between observations \cite{Li2024}.

The small changes in the aggregate metrics after ICME exclusion suggest that the cataloged intervals do not explain most of the full-year SWX-2D forecast error.
The low correlations during some ICME-rich Carrington rotations may reflect disturbances extending beyond the padded catalog intervals, but these descriptive associations do not establish a causal contribution from ICMEs.

As illustrated in Section~\ref{sec:2d}, the new interpolation introduces a variable-sign timing shift, changing not only the smoothness of the resulting time series but also the timing and shape of the streams observed at a spacecraft.

We assessed timing offsets using hourly series and $r(\tau)=\mathrm{corr}[v_{\mathrm{obs}}(t),v_{\mathrm{forecast}}(t+\tau)]$, scanning $\pm5$ days; negative lags indicate forecast structures occurring earlier than observed.
At ACE, all available observation--forecast pairs were used at each lag.
For SWX-2D with the empirical slow SW model, the correlation increased from $r(0)=0.53$ ($N=7314$) to $r_{\max}=0.55$ at $\tau=-8~\mathrm{h}$; for SWX, it increased from $r(0)=0.57$ ($N=3552$) to $r_{\max}=0.61$ at $\tau=-10~\mathrm{h}$.
Both forecasts thus showed optimum timing offsets of approximately ten hours on their respective available samples; their differing temporal coverage prevents interpreting the small difference as evidence of superior timing.

For the full-year STEREO\mbox{-}A analysis, each lag used common hours with an observation and recurrent forecast at $t$ and a SWX-2D prediction at $t+\tau$, without ICME exclusion.
The optimum was close to zero, $\tau=+4~\mathrm{h}$, increasing the correlation from $r(0)=0.398$ ($N=5628$) to $r_{\max}=0.405$ ($N=5634$).

During CR~2206, pairwise lag optimization increased the STEREO\mbox{-}A correlation from $0.15$ to $0.51$ at $\tau=-91~\mathrm{h}$; extending the lag window to $\pm10$ days moved the optimum to $-211~\mathrm{h}$.
With no unambiguous modeled counterpart to the dominant observed HSS, these large shifts align different streams and should not be interpreted as physical arrival-time corrections.

The upper tail of the SWX-2D speed distribution is systematically lower than that of the observations.
On paired model-observation timestamps, the 95th percentile of the modeled solar wind speed is 55 km/s below the observed value at ACE and 95 km/s below it at STEREO-A.
We attribute this effect primarily to the empirical CH area-SW speed mapping; repeating the calibration of this closure for the 25th solar cycle is a priority for future work.

Since SW speed in SWX-2D is inferred from coronal hole areas, factors that hinder the direct identification of their boundaries also directly affect SWX-2D performance.
Solar filaments, which consist of cool, dense plasma, appear as dark regions in extreme-ultraviolet images \cite{Chen2025}.
We examined the results of applying the threshold-based coronal hole detection algorithm to the COSPAR ISWAT catalog \cite{Reiss2024}, which contains solar images that are difficult to segment and is used to evaluate different coronal hole detection methods.
We found 15 instances in which filaments were identified as coronal holes, affecting 10 of the 28 images in the catalog.
Several approaches can be used to exclude filaments, including morphological criteria, SDO/HMI magnetograms in which coronal holes exhibit pronounced magnetic polarity \cite{Hofmeister2019}, and ground-based $H\alpha$ observations \cite{Joshi2010,Berezin2023}.

Because hydrodynamic interactions between SW parcels are not implemented in SWX-2D, it cannot reproduce the rarefaction and compression regions that arise during transitions between slow and fast solar wind.
Although SWX-2D does not predict solar wind density, approximate background density values could be assigned using empirical density-speed relationships, as employed in heliospheric models such as EUHFORIA and SURF-hydro \cite{Pomoell2018, Owens2026}.
However, the present ballistic propagation scheme does not conserve mass during stream overtaking, and therefore cannot reproduce the density enhancements associated with CIR compression regions.
Introducing mass conservation and a physically motivated treatment of stream interactions represents a possible direction for future development.

Slow solar wind, whose speed SWX-2D is not equipped to predict, constitutes most of the observed solar wind \cite{McGregor2011}.
Its variability may therefore dominate that of high-speed streams in metrics that do not treat high-speed streams as discrete events.
Validation against catalogs containing the time intervals during which high-speed streams were observed \cite{Yermolaev2009} could make the metrics more sensitive to high-speed stream variability \cite{Owens2005a}.
At the same time, however, it could introduce additional subjectivity because the identification of high-speed streams in these catalogs is itself subjective.

\section{Conclusion}

We have developed SWX-2D, a two-dimensional extension of the 1D CH area-based SWX forecast \cite{Shugai2021} that covers the heliosphere in the ecliptic plane.
Comparison of SWX-2D against SWX confirmed that the original forecast mechanism has been preserved after the transition to the new interpolation and propagation architecture.
When the empirical slow SW model is applied to SWX-2D at ACE, SWX-2D and SWX have a correlation of $r=0.93$ and an RMSE of $30~\mathrm{km\,s^{-1}}$.
Because the forecasts share their empirical components, this result represents a consistency test rather than an independent estimate of forecast skill.

Independent validation against in situ measurements showed that SWX-2D reproduced more of the observed variability at ACE than a recurrent forecast, with $r=0.55$ compared with $r=0.44$ for all data and $r=0.55$ compared with $r=0.46$ after ICME exclusion.
At STEREO\mbox{-}A, however, the recurrent forecast performed substantially better, with $r=0.63$ compared with $r=0.40$ for SWX-2D.
Excluding cataloged ICMEs did not materially improve SWX-2D at either spacecraft, indicating that unmodeled ICMEs were not the principal source of error during the studied interval.

Despite its lower accuracy at STEREO\mbox{-}A, SWX-2D retained some location\mbox{-}specific information away from the Earth--Sun line.
On the $N=5439$ timestamps common to all three series, the ACE--STEREO\mbox{-}A correlation was $r=0.28$ after shifting ACE by a fixed 19~days, whereas the SWX-2D--STEREO\mbox{-}A correlation was $r=0.39$.
The agreement between SWX-2D and STEREO\mbox{-}A was therefore stronger than the agreement between STEREO\mbox{-}A and the fixed-delay ACE measurements.
Its limited accuracy is instead consistent with the considerable age of the Earth-view observations used to predict solar wind at STEREO\mbox{-}A, which moved from approximately $120\degree$ to $100\degree$ west of Earth during 2018: 18--20~days at the start of propagation and 21--26~days at spacecraft arrival after solar wind propagation, as described in Section~\ref{sec:validation_setup}.

At ACE, both forecasts showed optimum timing offsets of approximately ten hours; differing temporal coverage prevents a direct comparison of their timing accuracy.

\printbibliography

\end{document}